# Nonlinear scalings emerge in a linear regime: an observation in electrokinetic flow

Jin'an Pang [1], Guangyin Jing [2], Xiaoqiang Feng [1], Kaige Wang [1,*], Wei Zhao [1,*]

[1] State Key Laboratory of Photon-Technology in Western China Energy, International Collaborative Center on Photoelectric Technology and Nano Functional Materials, Laboratory of Optoelectronic Technology of Shaanxi Province, Institute of Photonics & Photon Technology, Northwest University, Xi'an 710127, China;
[2] School of Physics, Northwest University, Xi'an 710127, China
* Correspondence: wangkg@nwu.edu.cn; zwbayern@nwu.edu.cn

In nonlinear systems, small perturbations are conventionally attributed to negligible nonlinearity, justifying linear approximations. Here, we uncover a notable exception to this paradigm in an electrokinetic (EK) flow. Using a novel dual-frequency excitation scheme with two high-frequency AC electric fields ($> 10^5$ Hz), we efficiently excite flow perturbations at a difference frequency ($\Delta f$) four orders of magnitude lower. This approach reveals a strong nonlocal energy transfer mechanism mediated purely by the nonlinearity of the electric body force, enabling precise, clean flow control free from electrode polarization artifacts. Unexpectedly, these small, nominally linear velocity and electric conductivity fluctuations exhibit power-law spectra. With increasing electric Rayleigh number, the scaling exponents agree quantitatively with predictions for fully developed EK turbulence by the Quad-cascade process theory. This observation not only implies multiple flow-state transitions even at low excitations, but also indicates that intrinsic nonlinearity regulates perturbations even in the linear regime, necessitating a fundamental re-examination of linear approximations in electrohydrodynamics and other nonlinear systems.

Perturbations are ubiquitous in nature, governing physical and biological processes, yet their mechanisms remain insufficiently understood in nonlinear systems. In fluid dynamics, flows can transition to chaos or turbulence via intrinsic instabilities or external forcing including buoyancy (1), electric body force (EBF) (2, 3), Lorentz force (4), or active matter (5, 6) etc. To understand the response of flow to perturbations, researchers have developed several theoretical tools, e.g. linear instability (7), weakly nonlinear instability (7) and nonmodal theory (8). However, even if the flow perturbation is negligibly small, it is difficult to predict the receptivity of flow accurately due to inherent nonlinearity. A major experimental bottleneck lies in the difficulty of injecting precise and clean perturbations noninvasively and locally on demand. In this investigation, via well-controlled EBF excitation in electrokinetic (EK) flow, we experimentally demonstrate that the flow receptivity exhibits scaling and nonlinear features, even if the flow perturbations are so small that the flow system is commonly regarded as linear.

Perturbations of EK flow can be precisely generated by EBF exerted on an interface with nonuniform electric conductivity ($\sigma$) (9, 10) (Fig. 1A, B), if an external AC electric field is applied (Fig. 1C). The flow velocity ($\boldsymbol{u}$) and $\sigma$ are governed by Navier-Stokes equation (11) with EBF ($\boldsymbol{F_e}$) shown in Eq. (1), and transport equation of $\sigma$ shown in Eq. (2) respectively,

$$\frac{\partial \boldsymbol{u}}{\partial t} + \boldsymbol{u} \cdot \nabla \boldsymbol{u} = -\frac{\nabla p}{\rho} + \nu \nabla^2 \boldsymbol{u} + \frac{\boldsymbol{F_e}}{\rho} \quad (1)$$

$$\frac{\partial \sigma}{\partial t} + \boldsymbol{u} \cdot \nabla \sigma = D_\sigma \nabla^2 \sigma \quad (2)$$

where $p$ is pressure, $\nu$ is kinematic viscosity, $D_\sigma$ is effective diffusivity of $\sigma$. Given the flow is incompressible in this low Reynolds number microchannel flow, $\nabla \cdot \boldsymbol{u} = 0$ and fluid density $\rho$ is constant. When the electric permittivity ($\varepsilon$) of fluids can be approximated to constant, the influence of dielectric becomes negligible, while Coulomb force is dominant (12). In the conditions that complex Ohm's law is established, the local electric field $\boldsymbol{E} = \boldsymbol{J}^*/\sigma^*$ (13), where $\boldsymbol{J}^*(t) = V\sigma_w^*/w$ is complex electric current density which can be approximated to spatially constant along the direction of external AC electric field, $V = V(t)$ is applied AC voltage, $\sigma^* = \sigma + i\omega\varepsilon$ is complex electric conductivity, $\omega$ is angular frequency and $\sigma_w^*$ is an reference $\sigma^*$. Thus, approximately we have (12)

$$\boldsymbol{F_e} \approx -\mathrm{Re}\left[\left(\varepsilon^* \boldsymbol{E} \cdot \frac{\nabla \sigma}{\sigma^*}\right) \boldsymbol{E}\right] \quad (3)$$

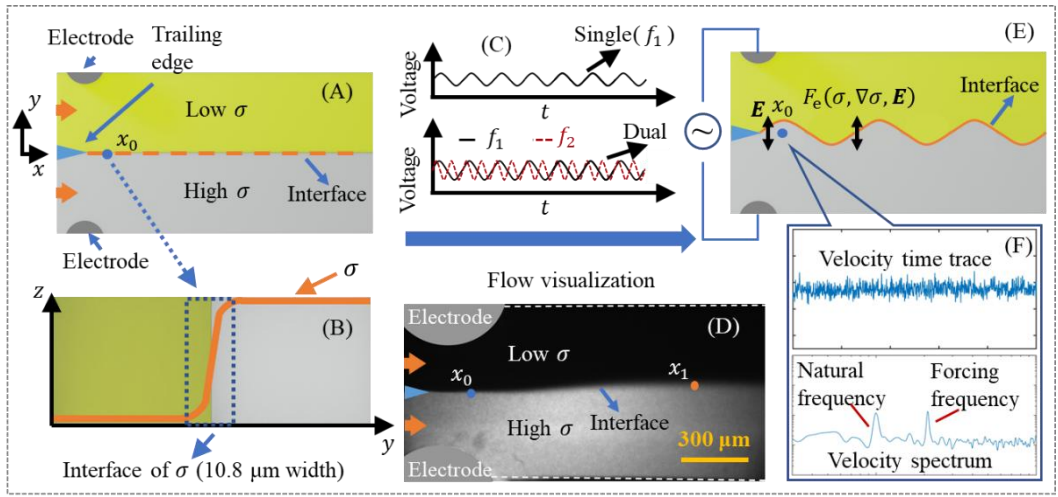

Fig. 1. Diagram of the investigation. A. Schematic of the microchannel flow without excitation. B. Spanwise distribution of $\sigma$ without excitation. C. Schematic of AC voltages in single- and dual-frequency excitations. D. Schematic of the EK flow with excitation. E. Flow visualization of unexcited flow. The blue spot illustrates the measurement point, which is on the interface and $x_0 = 150$ μm downstream of the trailing edge. $x_1 = 2$ mm is the position where velocity calibration is carried out. F. Velocity time trace and its spectrum.

Herein, $\boldsymbol{E}$ is coupled to $\sigma$, which in turn modulates $\boldsymbol{F_e}$ and $\boldsymbol{u}$, forming an internal relationship among $\sigma$, $\boldsymbol{u}$ and $\boldsymbol{E}$. It can be inferred from Eq. (3), a well-defined conductivity gradient under an applied electric field yields precisely controllable flow perturbation through $\boldsymbol{F_e}$ (Fig. 1D).

The experimental design is introduced in supplementary materials. Briefly, the experiments were conducted in the microfluidic chip shown in Fig. 1E and Fig. S1A. The chip is 18 mm long, with a minimum width of 650 μm at trailing edge and a uniform height of 100 μm. Solutions with an electric conductivity ratio of 1:5000 were injected at 5 μL/min from inlets 1 and 2 giving a bulk Reynolds number $Re_b = UL/\nu = 0.4$, where $L = 2dh/(d+h)$ is the hydraulic diameter of mixing chamber and $U$ is the bulk flow velocity. A dual-frequency AC voltage (two independent AC voltages with different frequency and phase, see Fig. 1C) was applied to electrodes to generate a y-directional EBF at the conductivity interface. The resulting tiny velocity fluctuation, $v'$ (e.g.

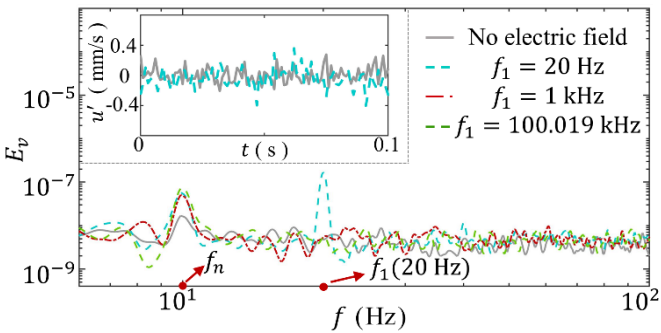

Fig. 2. Velocity spectra under the single-frequency excitation under $Q = 5$ μL/min with different frequency.

Fig. 1F), was measured at the interface center, 150 μm downstream of the trailing edge, using a laser-induced fluorescence photobleaching anemometer (LIFPA) system with ultra-high spatial (~180 nm) and temporal (~4 μs) resolution (14, 15). Electric conductivity fluctuations were measured in-situ via laser-induced fluorescence (LIF) with the same optical system at a sampling rate of 1 kHz. Time traces were analyzed in the spectral domain to characterize receptivity (Fig. 1F).

Without an electric field, the laminar wake exhibits small random velocity fluctuations, with a natural frequency $f_n \sim 10$ Hz at $Q = 5$ μL/min (Fig. 2). Applying a single high-frequency AC field (e.g., $f_1 = 100$ kHz) yields a single peak at $f_n$, while a low-frequency field (e.g., $f_1 = 20$ Hz) produces an additional peak at $f_1$, consistent with prior observations (16). No nonlinear peaks are observed under single-frequency excitation.

In contrast, applying a dual-frequency field with a low electric Rayleigh number ($Ra_e$) (9) triggers triad resonance via the second-order nonlinearity of the EBF. As shown in Fig. 3A, with $\Delta f = f_2 - f_1$ ($f_2$ being the second AC frequency) fixed at 7 Hz or 11 Hz and $f_1$ varied, clear peaks appear at $f_1$, $f_2$, $\Delta f$, and $f_1 + f_2 (= f_{sum})$. Remarkably, even when $f_1$ and $f_2$ are at 100 kHz (four orders higher than $\Delta f$) or above, a strong peak at $\Delta f$ persists (Fig. 3B). This demonstrates that the nonlinearity of the electric field in the EBF — not the fluid velocity through any transfer loop (17) —primarily controls this important nonlocal energy transfer. The nonlinear excitation is so effective that, dual-frequency excitation achieves a high efficiency up to 85% of single-frequency excitation (Fig. 3C), despite the vast frequency separation.

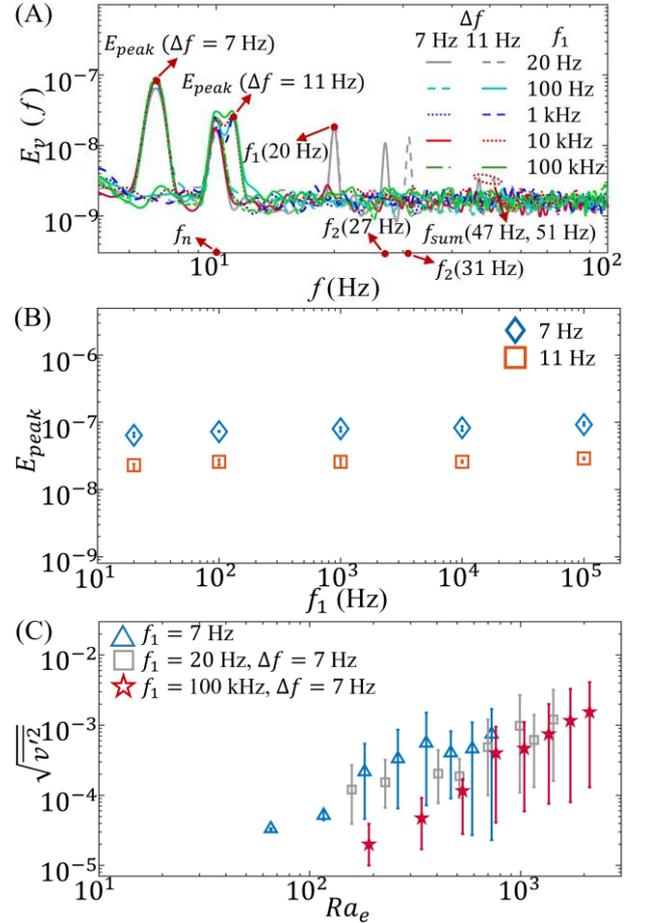

Fig. 3. Comparison between single- and dual-frequency excitation. A. Velocity power spectra when excited by dual frequencies under $Q = 5$ μL/min. Here, $\Delta f = 7$ Hz and 11 Hz, $Ra_e \approx 228.2$ with different $f_1$. B. Peak intensity of $E_v$ at $f = \Delta f$ vs $f_1$ after denoising. C. Receptivity evaluated by $\sqrt{\overline{v'^2}}$ in both single- and dual-frequency cases after denoising.

An additional advantage of dual-frequency excitation from high frequencies by nonlocal energy transfer, is its *cleanness*. It avoids the thick, oscillating electrode triple layers (ETL) (*18-20*) on electrodes which could generate nonlinear electric field during single-frequency excitation at a low $f_1$. Thus, a precise low-frequency flow modulation can be precisely achieved.

By maintaining a low AC voltage, we ensured that nonlinear flow dynamics were negligible; only the linear mode at $f = f_1$ (single-frequency) or $f = \Delta f$ (dual-frequency) was observed, without harmonic components. Despite this, the velocity fluctuation spectra ($E_v$) revealed surprising power-law scaling. As shown in Fig. 4A, B, the spectral peak energy $E_{peak}$ decays with frequency $f$ following $E_{peak} \sim f^{\beta_u}$. The exponent $\beta_u$ is not universal but varies with $Ra_e$ (Fig. 4C), showing four typical values, e.g. $-2.01 (\approx -2)$ at $Ra_e = 172.1$, $-1.71 (\approx -5/3)$ at $Ra_e = 248.6$, $-1.40 (= -7/5)$ at $Ra_e = 293.3$, and $-1.02 (\approx -1)$ at $Ra_e = 368.1$, respectively.

Similarly, the scalar spectra for electric conductivity fluctuations ($E_s$, see supplementary materials for details) exhibit scaling behavior $S_{peak} \sim f^{\beta_s}$ with exponents of $-3.00 (\approx -3)$ $-2.30 (\approx -7/3)$, $-1.88 (\approx -9/5)$, and $-1.69 (\approx -5/3)$ for $Ra_e = 172.1, 248.6, 293.3$, and $368.1$, respectively (Fig. 4D).

The observed scaling exponents (Table 1) are consistent with those predicted for fully-developed three-dimensional EK turbulence by the Quad-cascade process model (*21-23*). This model predicts, for instance, a $-2$ velocity spectrum and a $-3$ scalar spectrum in a variable-fluxes regime, a $-5/3$ velocity spectrum and a $-7/3$ scalar spectrum in a constant energy flux regime, and $-7/5$ and $-9/5$ in a constant scalar variance flux regime. All these scalings are exactly recurred at $Ra_e = 172.1, 248.6$ and $293.3$. They indicate an entanglement between $E_{peak}$ and $S_{peak}$ as $E_{peak}^2 f \sim S_{peak}$ in an EK flow, corresponding to a commonly used relationship that $F_e \sim E^2 \nabla \sigma$ (*23*). These scaling laws are hallmarks of highly nonlinear dynamics and are not expected to appear in a linearized flows with small perturbations where vortices are absent. Our findings therefore indicate that the intrinsic nonlinearity of the flow dynamics regulates perturbations even in regimes traditionally considered as linear. This necessitates a fundamental re-evaluation of the role of linear approximations in fluid mechanics and other nonlinear systems, from geophysics and astrophysics to condensed matter and biological physics (*24-28*).

Furthermore, at $Ra_e = 368.1$, we observe a kinetic energy spectrum with a slope of $\beta_u \approx -1$, indicating much slower spectral decay than expected. This behavior is atypical in related disciplines such as electrohydrodynamic and magnetohydrodynamic turbulence. In contrast, the scalar spectrum exhibits a slope of $\beta_s = -5/3$, suggesting scalar fluctuation governed by an inertial-subrange-like passive transport mechanism. In this case, $E_{peak}^2 f^{1/3} \sim S_{peak}$ which implies $F_e \sim E^2 \nabla^q \sigma$ with $0 < q < 1$ (*23*), which could be attributed to an unexpected intermedium state where anomalous diffusion (*29*) may exist.

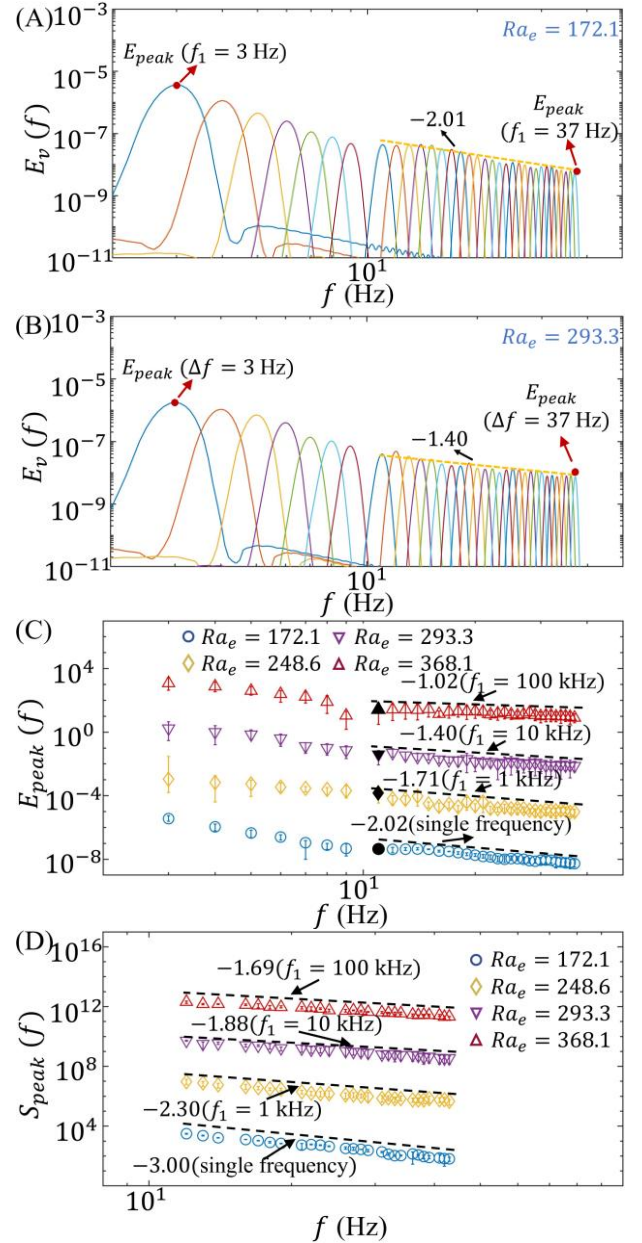

Fig. 4. Velocity power spectra at $Q = 5$ μL/min. A. Velocity power spectra of EK flow under single-frequency excitation $Ra_e = 172.1$ with different $f$. B. Velocity power spectra of EK flow under dual-frequency excitation under $Ra_e = 248.6$ and $f_1 = 1$ kHz with different $\Delta f$. C. $E_{peak}$ vs $f$. All $\beta_u$ are calculated by nonlinear fitting on the averaged $E_{peak}$ over 3 datasets. D. $S_{peak}$ vs $\Delta f$ under EBF excitation at $Q = 5$ μL/min. All $\beta_s$ are calculated by nonlinear fitting on the averaged $S_{peak}$ over 3 datasets.

Table 1. Summary of the scaling exponents vs $Ra_e$

| $Ra_e$ | 172.1 | 248.6 | 293.3 | 368.1 |
|---|---|---|---|---|
| $\beta_u$ | -2 | -5/3 | -7/5 | -1 |
| $\beta_s$ | -3 | -7/3 | -9/5 | -5/3 |

The multiple scaling behaviors observed with increasing $Ra_e$ carry far-reaching implications across diverse branches of physics. In fluid dynamics and condensed matter physics, even at low $Ra_e$ where velocity nonlinearity remains negligibly small, the electrokinetic flow already undergoes a sequence of *transitions through multiple intermediate states*. For nonlinear dynamical systems, these observations provide direct experimental evidence that self-similarity can emerge and persist within non-critical (regular) regimes (*30*), rather than arising solely at critical points. Rather than a single fixed scaling relation, a branch of *self-similar bridges* may evolve continuously toward the critical region.

The mechanism of *nonlinear nonlocal energy transfer uncovered in the linear regime* possesses universal, scale-spanning significance. It extends well beyond classical electromagnetism and fluid dynamics, offering new conceptual insights into quantum physics and condensed matter systems. For example, in the actuation of piezoelectric and dielectric elastomer materials, dual-frequency excitation provides a highly efficient and precise route to generate low-frequency mechanical forcing. In quantum electrodynamics, both Schwinger effect (*31*) and Breit-Wheeler process (*32*) are representative predictions. The possibility of generating electron-positron pair, even in a linear approximation under low electromagnetic field, may still be regulated by the intrinsic nonlinearity of quantum physics system. Similar results can also be spanned to quantum fluids.

Additionally, the dependence of phenomena such as quantum dot transport (*33*) and Lamb shifts (*34, 35*) in mesoscopic quantum systems on nonlocal interactions echoes the power spectrum scaling law of this study, indicating that nonlinear regulation is a common logic underlying quantum fluctuation. Therefore, probing the spectrum scaling law of tiny electric signal could be an effective approach for revealing the nonlinear interaction between vacuum fluctuations and matter.

In this investigation, we have reported two principal findings. First, we demonstrated a novel nonlocal energy transfer where the nonlinearity of the electric body force enables efficient excitation from high-frequency AC fields ($\geq 10^5$ Hz) to excite flow perturbations at a frequency four orders lower. This provides a clean and powerful tool for precise flow control. Second, and more profoundly, we showed that these small, linear-flow perturbations exhibit power-law scaling identical to that of fully developed EK turbulence. This reveals that nonlinearity is an inherent regulator of flow dynamics, even in the linear regime, challenging the universality of linear approximations.

Our work not only advances the understanding of electrokinetic turbulence but also provides a new perspective for investigating nonlinearity across a wide range of physical systems. These cross-disciplinary mappings not only blur the boundary between macroscopic fluids and microscopic quantum systems, but also provide a horizon of *verifying high-energy physics theories through condensed matter systems*. Through nonlinear electromagnetic interaction and nonlocal energy transfer as a unifying language, it is possible of connecting among isolated phenomena like turbulent cascades, particle creation, and energy level fluctuation, with macroscopic experimental supports.

**Acknowledgement** This research was supported by the National Natural Science Foundation of China (No. 62275216, 61775181), the Natural Science Basic Research Program of Shaanxi Province-Major Basic Research Project (No. 2025SYS-SYSZD-003). We also appreciate the instructive suggestions from Jianzhou Zhu.